\documentclass[aps,showpacs,nofootinbib]{revtex4}

\usepackage{graphicx}
\usepackage{graphics}
\usepackage{amssymb}
\usepackage{amsmath}
\usepackage{bm}

\newcommand{\be}{\begin{equation}}
\newcommand{\ee}{\end{equation}}
\newcommand{\bea}{\begin{eqnarray}}
\newcommand{\eea}{\end{eqnarray}}
\newcommand{\beal}{\begin{align}}
\newcommand{\eal}{\end{align}}
\newcommand{\bespl}{\begin{split}}
\newcommand{\espl}{\end{split}}

\newcommand{\nslash}{\kern 0.2 em n\kern -0.50em /}
\newcommand{\kslash}{\kern 0.2 em k\kern -0.45em /}
\newcommand{\pslash}{\kern 0.2 em p\kern -0.50em /}
\newcommand{\Sslash}{\kern 0.2 em S\kern -0.50em /}
\newcommand{\Pslash}{\kern 0.2 em P\kern -0.50em /}
\newcommand{\Rslash}{\kern 0.2 em R\kern -0.50em /}

\begin{document}

\title{A study on 
the feasibility of a precise measurement of 
the $\tau$-dependence of the cross sections for 
Drell-Yan experiments at moderate energies
}

\author{A.~Bianconi}
\email{andrea.bianconi@bs.infn.it}
\affiliation{Dipartimento di Chimica e Fisica per l'Ingegneria e per i 
Materiali, Universit\`a di Brescia, I-25123 Brescia, Italy, and\\
Istituto Nazionale di Fisica Nucleare, Sezione di Pavia, I-27100 Pavia, Italy}

\begin{abstract}
Recently, a reconsideration of Drell-Yan cross sections at 
moderate energies and masses has suggested the possibility of 
relevant enhancements of the cross sections in some kinematical 
regions. If confirmed, these predictions could largely affect the 
planning of Drell-Yan experiments aimed at transverse spin measurements 
after 2010. More in general, 
the problem is present of a precision measurement of the 
$\tau$ dependence of Drell-Yan cross sections. Here we discuss 
the feasibility of such a measurement 
within short time at the COMPASS apparatus, and its relevance 
for the PANDA experiment. 

\end{abstract}


\maketitle

\section{Introduction} 

In a recent series of papers\cite{BRMC} detailed Montecarlo studies 
have been produced for Drell-Yan 
$\mu^+\mu^-$ experiments at moderate
center of mass energies and dilepton masses ($S$ $\sim$ 30-400 GeV$^2$, 
$M$ $<$ 12 GeV/c$^2$). These experiments would measure 
unpolarized, single and double spin asymmetries 
and include proposals and plans at GSI\cite{panda,assia,pax},  
at RHIC\cite{rhic2}, and at COMPASS\cite{compass-hadron}. 

For the following discussion the relevant variables are 
the squared center of mass energy $S$, the parton longitudinal 
fractions $x_1$ and $x_2$, the dilepton mass $M$, and 
$\tau$ $\equiv$ $M^2/S$ $=$ $x_1x_2$. 
At magnitude level 

\begin{equation}
{{d\sigma} \over {dx_1dx_2}}\ 
\sim\ N {{f(x_1) f(x_2)} \over {S \tau}}. 
\end{equation}
\noindent 
For the cases considered here 
$N$ can be 1 nbarn or smaller. The product $f(x_1)f(x_2)$ 
is a symbolic way to represent a bilinear combination of partonic 
distribution 
functions, that decreases fastly towards zero for $x_1$, $x_2$ 
or $\tau$ $>>$ 0.1. 

According to the above equation we have 
small overall event rates, 
fastly decreasing at increasing $\tau$;
However, all transverse spin related 
measurements require one or both of the longitudinal momentum 
fractions in the valence region, corresponding to average $\tau$ 
values that cannot be too small. 

In a recent publication\cite{SSVY} the cross sections for 
Drell-Yan 
$\mu^+\mu^-$ production in the kinematical ranges relevant for the 
quoted experiments has been reconsidered, 
with resummation of soft gluon emissions near the partonic 
threshold $\tau$ $=$ 1, 
rising the cross sections in a $\tau-$dependent way.   
The effect is predicted to be huge for $\tau$ $\sim$ 1, 
but present anyway, and sometimes strong, also for $\tau$ 
$<<$ 1. 
If compared to the numbers by \cite{BRMC}, the enhancement 
factors 
can range from 2 to over 10 in regions that are relevant for  
the quoted experimental proposals at $S$ $<$ 300 GeV$^2$ (for $S$ 
$=$ 300-500 GeV$^2$ the cross sections are fixed by 
experimental data, see e.g.\cite{conway,anass}). 

If the predictions of ref.\cite{SSVY} are respected, 
several plans 
for quantities to be measured at GSI could be reconsidered. 
This makes an experimental confirmation 
of the prediction, if possible, an urgent matter. Apart for 
competing predictions, it would be useful to 
establish the $\tau$-dependence of the cross sections as precisely 
as possible. 

In our opinion this confirmation is 
possible within a few years with the COMPASS hadronic beam 
facility\cite{compass-hadron}, using  
a pion beam at 50 GeV/c. 
The measurement is rather tricky. 
The predicted cross section enhancements 
can arrive to a factor 2, but at relatively large $\tau$, 
where events are few. When the  
cross sections are integrated over a large mass range (equivalently, 
$\tau$ range) these enhancements are much less evident. 

Here a Montecarlo simulation is organized to study the proposed 
measurement at COMPASS, and also the case of the 
PANDA experiment\cite{panda} at GSI, where the effect is expected 
to be especially strong. The Montecarlo 
apparatus is the same as described in ref.\cite{BRMC}, so all the related 
details can be found in that group of references. 
K-factors 
(i.e. overall $\tau-$dependent factors renormalizing the cross sections)
have been rewritten 
so that cross section values follow alternatively the  
behavior predicted in refs.\cite{BRMC} and \cite{SSVY}. 

\section{The simulation} 

For COMPASS e consider (negative) pion-nucleon Drell-Yan with  
$S$ $=$ 100 GeV$^2$. We assume that the 
measurement is performed on a fixed target with $Z/A$ $=$ 0.5. 
We produce 50,000 simulated events for 4 $<$ M $<$ 9 GeV/c$^2$, 
$x_F$ $>$ 0 and 1 $<$ $P_T$ $<$ 3 GeV/c. 
This means about 35,000 events in the 
lower mass bin 4-5 GeV/c$^2$. 
The cut $x_F$ $>$ 0 is related with the strong possibility that fixed 
target experiments 
have reduced acceptance for negative $x_F$ (see e.g. the event 
scatter plot in ref.\cite{conway}). 
The $P_T$ cutoffs are 
related with the selection of Drell-Yan events in strict sense 
(see the discussions in ref.\cite{BRMC}). 
We perform the simulation both for the case of ``normal'' cross 
sections, and with the 
enhancement factors shown in ref.\cite{SSVY}. So we will speak 
of ``enhanced'' and ``non enhanced'' data sets. 

Defining $\Delta M$ as the overall mass range, we 
divide it into 5 equal subranges, and compare the simulated 
event population of the larger mass bins with the population of the 
lowest mass one. Since the populations of the lower mass bins are 
practically equal in the enhanced and non enhanced cases 
(about 35,000 events), we may limit ourselves to studying the 
populations of the upper mass bins. The underying 
necessary assumption is that these populations are extracted from 
well normalized samples of 50,000 events. 

For each case (enhanced and non-enhanced 
cross sections) 
the simulation is repeated ten times, so that for each 
mass bin we may calculate the average number of events and its 
fluctuation. The results are shown in 
fig.1. In figs. 2 and 3  
we show the laboratory angle and energy distributions, useful 
for the discussion on the feasibility of the measurement. 

\begin{figure}[h]
\centering
\includegraphics[width=9cm]{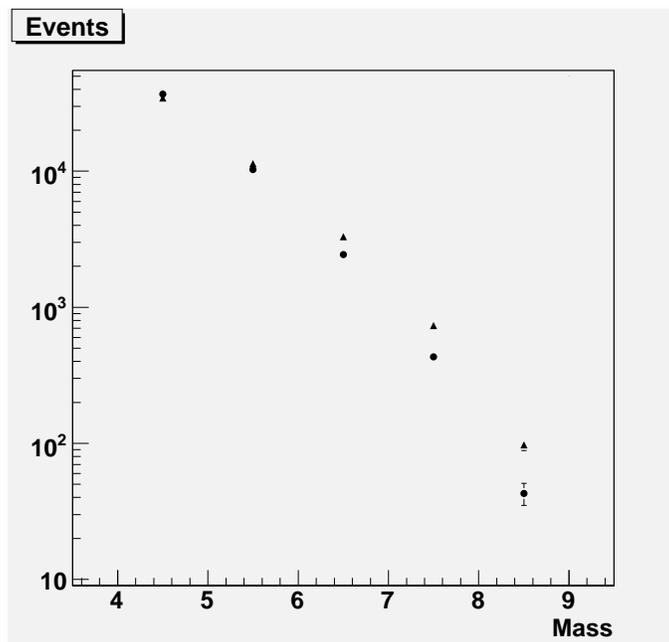}
\caption{
COMPASS: Event mass distribution for enhanced (triangles) 
and non-enhanced (circles) cross sections;
\label{compass}}
\end{figure}

\begin{figure}[h]
\centering
\includegraphics[width=9cm]{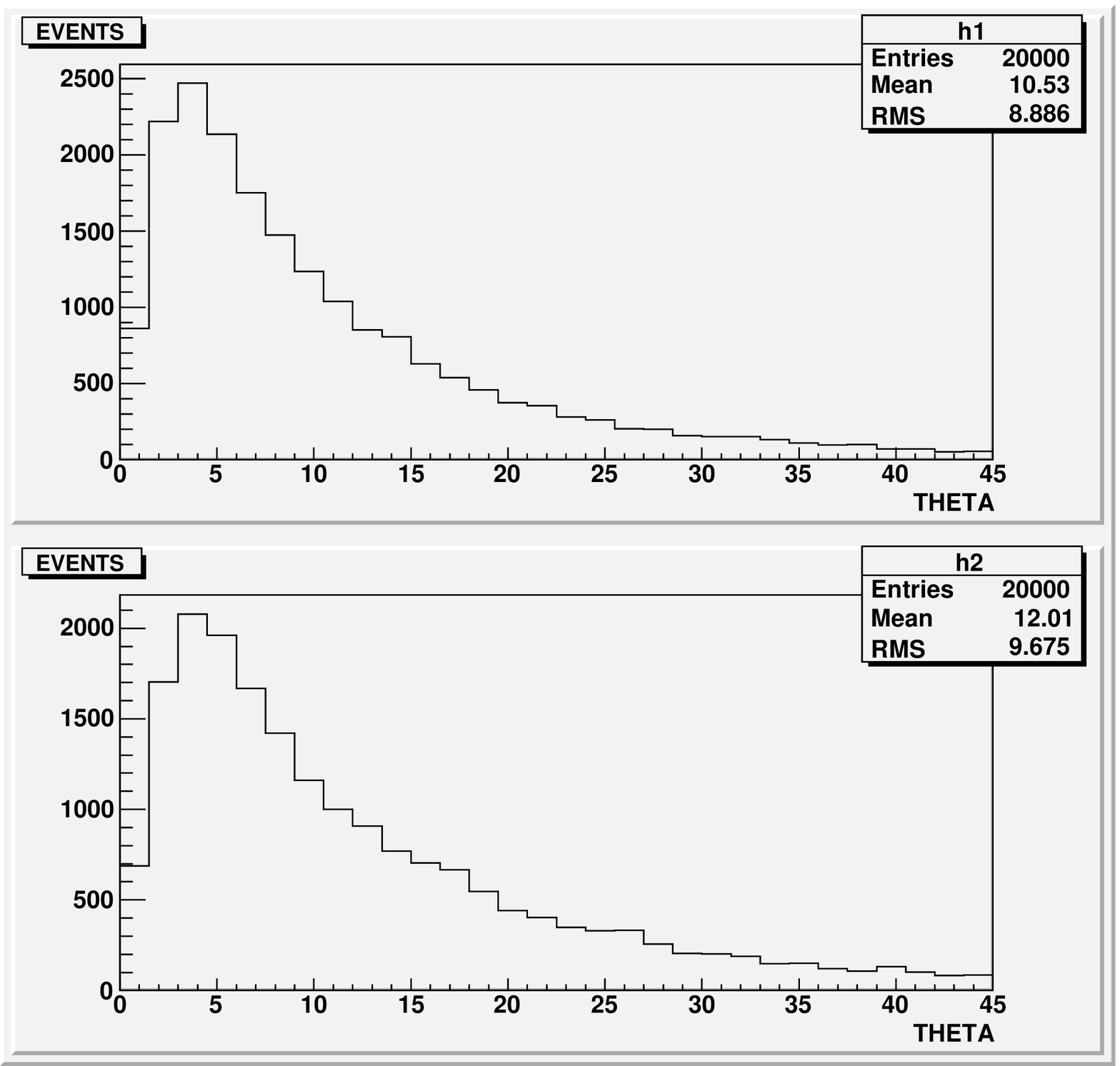}
\caption{
COMPASS: Upper panel: 
Laboratory polar angle distribution for (mixed) positive 
and negative muons for pair invariant mass in the range 4-5 GeV/c$^2$. 
Lower panel: the same for mass in the range 7-8 GeV7c$^2$. 
\label{theta_compass}}
\end{figure}

\begin{figure}[h]
\centering
\includegraphics[width=9cm]{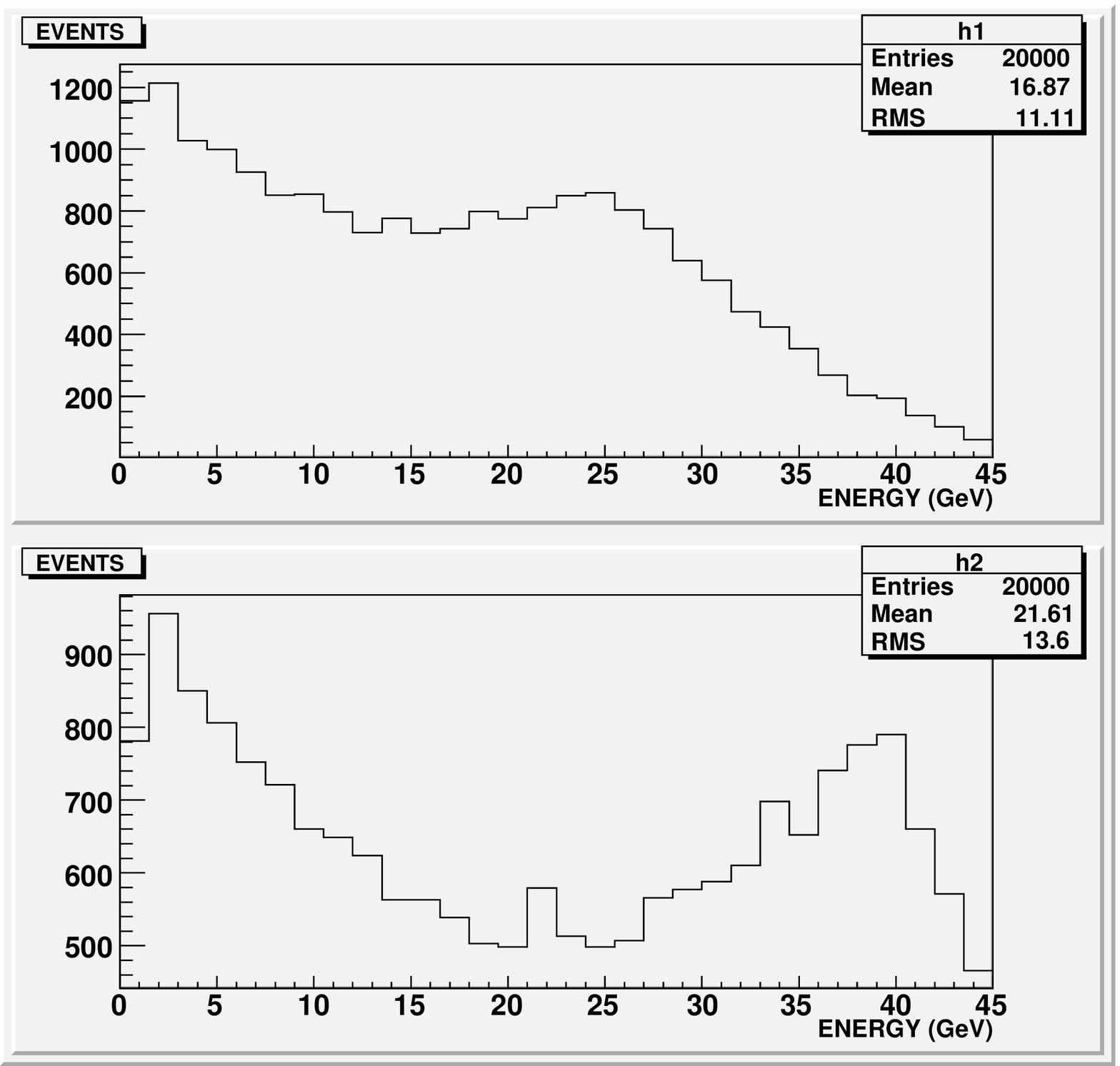}
\caption{
COMPASS: Upper panel: 
Laboratory energy distribution for (mixed) positive 
and negative muons for pair invariant mass in the range 4-5 GeV/c$^2$. 
Lower panel: the same for mass in the range 7-8 GeV7c$^2$. 
\label{energy_compass}}
\end{figure}

\begin{figure}[h]
\centering
\includegraphics[width=9cm]{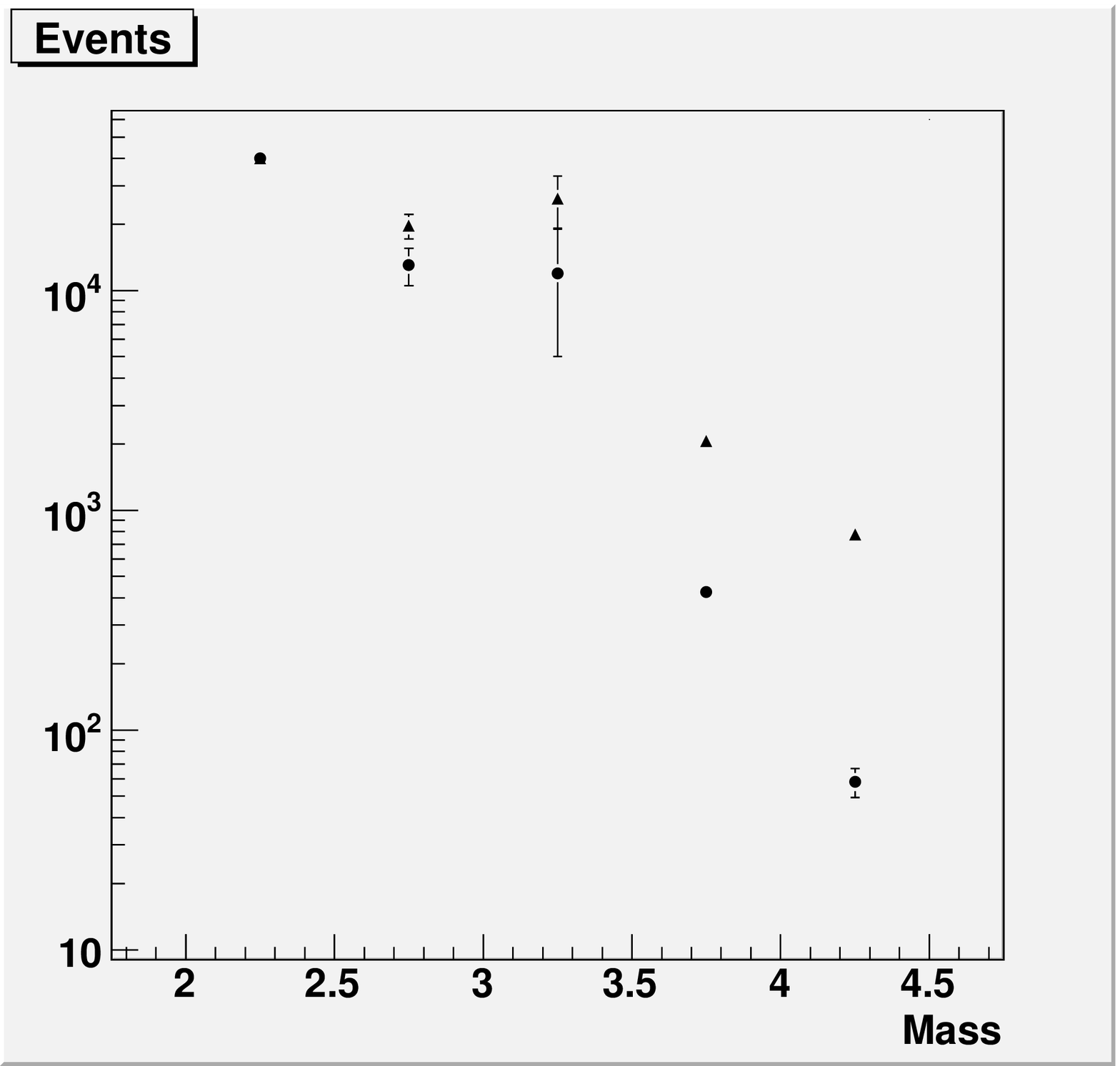}
\caption{
PANDA case: event mass distribution for enhanced (triangles) 
and non-enhanced (circles) cross sections; The large error bars 
in the second and third point are related with uncertainties 
in the prediction of the $J/\psi$ production rate. 
\label{panda}}
\end{figure} 

Fig.1 shows a 
delicate but interesting situation. In the last three mass 
ranges (6-7, 7-8, 8-9 GeV/c$^2$) the average event numbers are, 
in the enhanced vs non-enhanced cases: 

3300$\pm$60 vs 2443$\pm$57  

734$\pm$20 vs 432$\pm$19 

97$\pm$9 vs 43$\pm$8  

The statistical fluctuation is small enough to separate clearly the 
two possible outcomes in each mass bin. The very last 
mass bin contains a too small event number, so we forget about it and 
focus the rest of our analysis on the mass bin 7-8 GeV/c$^2$. 
The conclusions are valid for the mass bin 6-7 GeV/c$^2$ as well. 

The errors in the evaluation of the relative populations of the 
higher mass bin (7-8 GeV/c$^2$) to the 
lower mass bins (4-5 GeV/c$^2$) 
may arise from two sources especially: 

1) The background of random coincidence events with large mass, 
that in the 
case of a pion beam is due to coincidence between a 
beam halo muon with large energy and a collision-originated 
muon with small energy. 
The estimation of the fraction of ``wrong'' pairs 
in real experiments is normally based 
on the examination of like-charge pairs. According to the analysis 
in ref.\cite{conway}($S$ $\approx$ 470 GeV$^2$), and 
ref.\cite{anass}($S$ $\approx$ 250 GeV$^2$), this noise is not a 
problem 
since (i) the background falls as steeply as the true data frequency 
at increasing masses, (i) it remains at least one order of magnitude 
below data. 

2) The precision by 
which the relative acceptances for large and small mass events  
is estimated. This is potentially a problem if muons 
associated with large and small mass 
events distribute very differently 
in the laboratory frame phase space. 

Section 4 of ref.\cite{anass} is devoted to a  
detailed study of the acceptance, for a 
$\pi^-$-nucleus Drell-Yan experiment that presents 
some analogies 
with the COMPASS case. They show 
a flat efficiency in dilepton-track reconstruction for mass up 
to 6.8 GeV/c$^2$. The case of larger masses is not shown. 

In fig.2 we show 
the distribution of single muon polar angles in the 
(fixed target) laboratory reference frame.  
These muons belong to simulated Drell-Yan 
events in the above Compass conditions  
(without distinguishing between positive and negative muons). 
The upper panel reports the distribution of 10,000 events belonging to 
the dilepton mass region 
4-5 GeV/c$^2$, the lower panel an equal number of events belonging 
to the mass region 7-8 GeV/c$^2$. 
Despite there are differences, the two distributions are similar enough 
to exclude problems related with differential angular acceptance. 

In fig.3 we show the corresponding distributions for the muon 
laboratory energy. In this case there are important differences, 
but the major point is that both distributions assume 
non-negligible values in a large fraction of 
all the available energy phase space. 
An examination of the $\mu^+\mu^-$ correlations shows that for each event 
$E_+ + E_-$ $\approx$ constant (34 $\pm$ 7 GeV in 
the lower mass range, 45 $\pm$ 4 GeV in the higher mass range)
with a strong level of energy asymmetry 
$A_E$ $\equiv$ 
$2 \vert E_+ - E_- \vert / ( E_+ + E_- )$. In both cases 
$A_E$ follows an approximate distribution $1 + 1.4 (A_E)^2$ 
meaning that most events concentrate towards larger energy asymmetry. 

The above analysis means that identifying a dilepton pair requires, 
at small as at large masses, 
a good and well understood acceptance level throughout all the energy range, 
since the typical pair includes both a low-energy and a high-energy 
muon. A lack in this sense can decrease seriously the acceptance, 
but does not necesarily discriminate pairs 
with different masses. 

To better estimates the potential errors introduced by ignorance 
about energy acceptance, we simulate a really 
extreme situation: we suppose that the acceptance for single muons 
with energy $<$ 20 GeV 
is reduced by 50 \% and one is not aware of this. In other words, one is 
convinced that the apparatus acceptance is $\approx$ 1 at all energies, 
but this is true for $E$ $>$ 20 GeV only. 
In this case the fraction of silently lost pairs is 

49 \% in the low mass range, 46 \% in the high mass range. 

Things are worse in the opposite case (the acceptance is 
reduced to 50 \% for energy $>$ 20 GeV, and one is not aware of this). 
Now the fraction of lost pairs is 

40 \% in the low mass case, 51 \% in the high mass case. 

So, the lack of knowledge about true acceptance would lead to an 
error 20 \% in the estimation of the relative population of the 
two mass bins. 
The proposed examples are really pessimistic ones, since normally 
acceptance is reconstructed with far larger precision than this. 
Despite this, also in these situations the event ratio reconstruction 
is precise enough to estimate the searched enhancement factor. 

For PANDA (GSI) we 
consider $\bar{p}p$ Drell-Yan with 
$S$ $=$ 30 GeV$^2$, in the mass range 2-4.5 GeV$^2$. 
The other cuts are the same as for the COMPASS case. 
To avoid influence by the $J/\psi$ region on the event normalization,  
we normalize the collected data set 
with the requirement: 40,000 events 
in the mass range 2-2.5 GeV/c$^2$. 
For masses between 2.5 and 3.5 GeV/c$^2$ 
we base the simulation 
on Drell-Yan data by ref.\cite{biino} at large $x_F$, 
where gluon-gluon $J/\psi$ production is partially 
suppressed as it would happen at PANDA. 
This still leaves room for a large uncertainty factor in the 
$J/\psi$ production rates (see fig.4). 
We assume the threshold enhancement to be  
the same for the $J/\psi$ peak and for the background. 
As evident from fig.4, for masses over 3.5 GeV/c$^2$ event numbers 
increase by one order and 
there is no doubt on 
the benefit PANDA would receive from the enhancement. 

To summarize, the 
threshold enhancement can be verified 
with satisfactory precision 
in the COMPASS apparatus for pion beams. 
The enhancement would be much more striking 
at PANDA (GSI), 
with large increases in the counting rates at masses 
$>$ 3.5 GeV/c$^2$.



\end{document}